\documentclass[aps,prl,floatfix,twocolumn]{revtex4}

\usepackage{amsmath}
\usepackage{amssymb}
\usepackage{graphicx}
\usepackage{epstopdf}
\usepackage[usenames]{color}
\usepackage{cancel}
\usepackage{multirow}

\begin{document}

\newcommand{\bra}[1]{\mbox{$\langle #1|$}}
\newcommand{\ket}[1]{\mbox{$|#1\rangle$}}
\newcommand{\comment}[1]{\textit{\small\textcolor{red}{ #1}}}

\title{Polarization-entanglement conserving frequency conversion of photons}

\author{S.~Ramelow$^{1,2}$, A.~Fedrizzi$^{2,*}$, A. Poppe$^{1,3}$, N.~K.~Langford$^{1,2,\dagger}$, and A.~Zeilinger$^{1,2}$}

\affiliation{
$^1$Vienna Center for Quantum Science and Technology (VCQ), Faculty of Physics, University of Vienna, Boltzmanngasse 5, A-1090 Vienna, Austria \\
$^2$Institute for Quantum Optics and Quantum Information, Austrian Academy of Sciences, Boltzmanngasse 3, A-1090 Vienna, Austria \\
$^3$Austrian Institute of Technology (AIT), Donau-City-Stra{\ss}e 1, 1220 Vienna, Austria}

\begin{abstract}
Entangled photons play a pivotal role in the distribution of quantum information in quantum networks. However, the frequency bands for optimal transmission and storage of photons are not necessarily the same. Here we experimentally demonstrate the coherent frequency conversion of  photons entangled in their polarization, a widely used degree of freedom in photonic quantum information processing. We verify the successful entanglement conversion by violating a Clauser-Horne-Shimony-Holt (CHSH) Bell inequality and fully confirm that our characterised fidelity of entanglement transfer is close to unity using both state and process tomography. Our implementation is robust and flexible, making it a practical building block for future quantum networks.
\end{abstract}

\maketitle

Quantum frequency conversion of single photons offers an elegant way to avoid the often difficult trade-offs of choosing a single optimal photon wavelength. Quantum networks~\cite{duan_long-distance_2001} that will for example facilitate the large-scale deployment of secure quantum communication~\cite{scarani_security_2009} require the distribution of entanglement using flying qubits (photons) between quantum repeater nodes which can coherently store entanglement in quantum memories and concatenate it by entanglement swapping~\cite{pan_experimental_1998}. The standard wavelength for optical fiber transmission is $1550$~nm, where loss is minimal, whereas the highest quantum memory efficiencies have to date been achieved in Rubidium vapour at around 800 nm~\cite{hosseini_high_2011}. Such issues also arise in many other contexts, connected to e.g. detector performance (for 1550 nm photons), general transmission and dispersion properties of the materials used or the availability of suitable laser sources. Coherent frequency conversion of flying qubits can sidesteps this type of problem altogether.

The basic principle of optical frequency conversion is the nonlinear process of sum frequency generation (SFG): a pump and an input field are combined in a nonlinear medium generating an output field with the sum of the input frequencies. One main experimental motivation has been to solve the detection problem for single photons at 1550 nm, by converting to the visible regime and using -- instead of InGaAs-based -- better performing Silicon-based photon detectors~\cite{albota2004esp,vandevender2004hes,langrock2005hes,thew2006lju}. In the single-photon regime, where the input field is much weaker than the pump, near-$100$\% conversion efficiencies can be achieved by optimising interaction nonlinearities (e.g., by using waveguides~\cite{langrock2005hes, thew2006lju}) and high pump field intensities (e.g. with cavities~\cite{albota2004esp}, or pulsed pump lasers~\cite{vandevender2004hes}). Critically, the SFG process can also conserve the quantum properties of the input light~\cite{kumar_quantum_1990} and fulfil several fundamental requirements for universal photonic quantum interfaces: firstly, the conversion process must preserve the photons' indistinguishability~\cite{takesue2008edu,odonnell2009tru} as well as their single photon character~\cite{rakher_quantum_2010}; secondly, it must also preserve quantum information, and in particular entanglement, stored in the photons. Especially, polarization entanglement is being very widely used in a plethora of quantum optical experiments because of its remarkable practical advantages of precise and easy generation, control and measurement. While phase- and polarization-maintaining conversion has been shown for classical fields~\cite{giorgi_frequency_2003,albota_polarization-independent_2006,vandevender_quantum_2007}, conservation of entanglement has so far been reported for time-bin-entangled photons~\cite{tanzilli2005pqi}. 

Here, we demonstrate coherent conversion of \emph{polarization-entangled} photons from $810$~nm to $532$~nm using a continuous-wave pump laser at 1550 nm. By violating a \emph{Clauser-Horne-Shimony-Holt} (CHSH) \cite{clauser1969pet} Bell inequality we stringently verified the entanglement transfer. Our results extend coherent single-photon frequency conversion to the widely used and therefore important degree of freedom of photon polarization.

\begin{figure}
\includegraphics[width=\columnwidth]{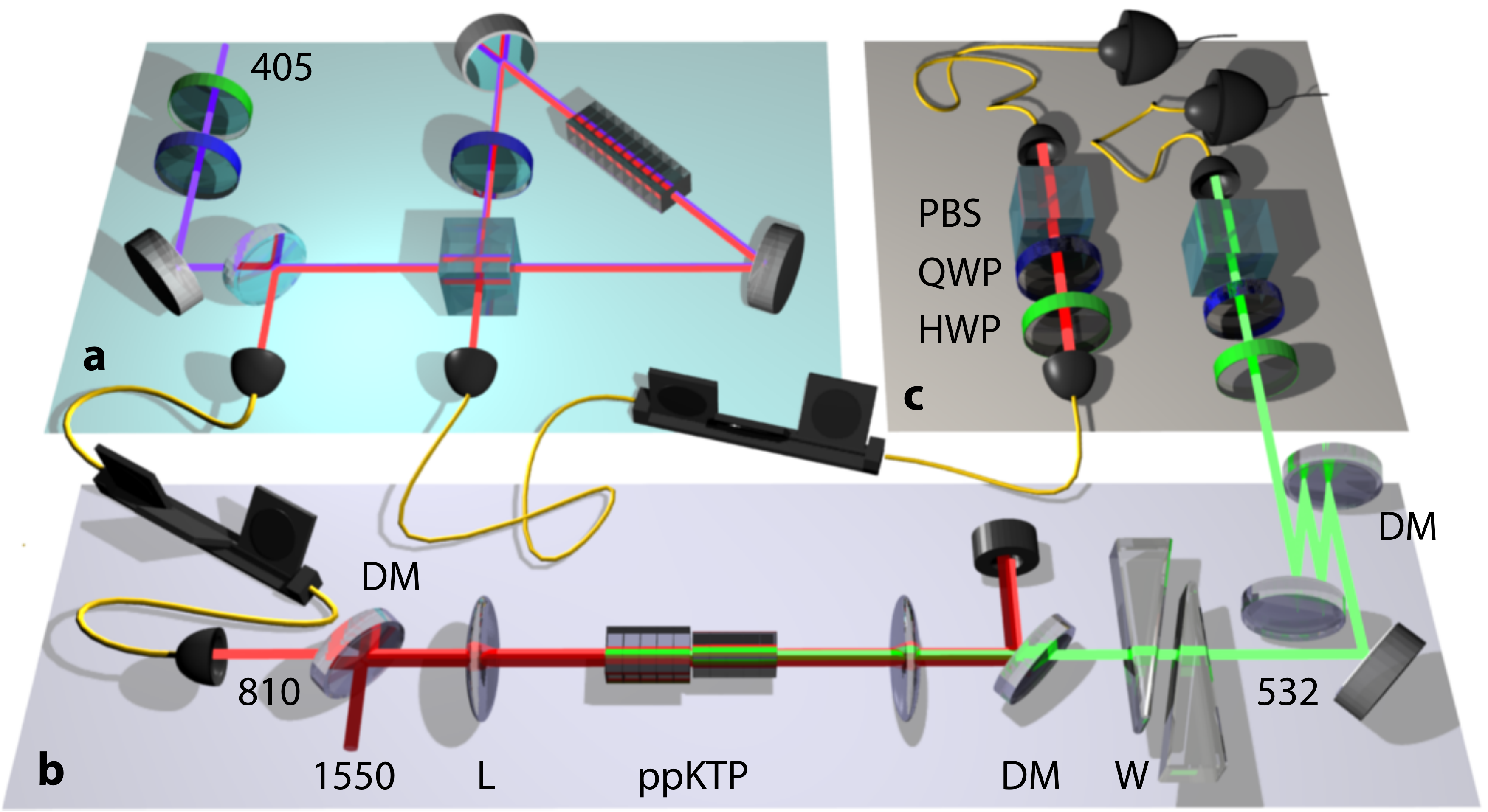}
\caption{Experimental scheme. (a) Polarization-entangled photon source. Photon pairs are created by spontaneous parametric downconversion in a periodically poled KTiOPO$_{4}$ (ppKTP) crystal which is bi-directionally pumped by a $405$~nm diode laser in a polarization-Sagnac loop \cite{kim_phase-stable_2006, fedrizzi2007wtf}. (b) Polarization coherent up-conversion setup. Signal ($810$~nm) and pump ($1550$~nm), are combined with a dichroic mirror (DM) and focussed into the setup with a $f=50$~mm lens (L). The two polarization components of the input are up-converted to $532$~nm in one of two $4.3$~mm long, orthogonally oriented ppKTP crystals. The remaining $810$~nm photons and $1550$~nm pump light are separated from the $532$~nm photons with dichroic mirrors: one in transmission and two in a z-configuration for multiple reflections. Two adjustable calcite wedges (W) compensate temporal walk-off. c) The polarization analysis (and preparation for the process tomography) is implemented with a quarter-wave plate (QWP), half wave plate (HWP) and a polarizing element (PBS). The $532$~nm photons are then coupled into a single-mode fiber and detected by a silicon avalanche photo-diode (Si-APD). The coincidences with the second $810$~nm photon are identified by home-built coincidence logic.}
\label{PolSpucSetup}
\end{figure}

In our experiment, figure~\ref{PolSpucSetup}, the polarization-coherent up-conversion takes place in two orthogonally oriented, periodically-poled  KTiOPO$_{4}$ (ppKTP) crystals. The crystals are designed for type-I quasi-phase-matching for 810~nm + 1550~nm $\rightarrow$ 532~nm, and oriented such that the horizontally (H) polarized component of the input at 810 nm is converted to $532$~nm (also H) in the first crystal, and the vertical (V) component is converted in the second. Chromatic dispersion and crystal birefringence cause a combined temporal walkoff of $\sim1.8$~ps between the orthogonal polarization components, similar to the photon coherence time~\cite{fedrizzi2007wtf}. To render the output photons indistinguishable, we compensate this walk-off with a pair of birefringent $CaCO_{3}$ (calcite) wedges with a combined thickness of $\sim3$~mm.
Thus, an input state $\phi^+_{\rm in}=(\ket{H_{810}H_{810}}+\ket{V_{810}V_{810}})/\sqrt{2}$ is converted into:
\begin{equation}\label{eq:conversion}
\psi_{\rm out}=\eta_{H}\ket{H_{810}H_{532}}+e^{-i\theta}\eta_{V}\ket{V_{810}V_{532}}.
\end{equation}
The phase $\theta$, as well as the relative conversion efficiency $\eta_{H}/\eta_{V}$ between the two crystals can be controlled through the polarization state of the $1550$~nm pump laser beam, which we adjust with fiber-polarization controllers. The pump laser system consists of a tunable, fiber-coupled external-cavity diode laser amplified to $1$~W with a high-power erbium-doped fiber amplifier. The pump field and the entangled photons were combined with a dichroic mirror and focused to spot sizes of $\sim70~\mu$m ($1550$) and $\sim50~\mu$m ($810$). After recollimation, the $532$~nm light was separated from both the 1550 nm pump and the 810 nm photons via multiple reflections off three dichroic mirrors, suppressing the unconverted 810 nm photons by at least 100 dB. The up-converted $532$~nm photon and its entangled $810$~nm partner photon were then subjected to polarization analysis and detected by single-photon avalanche photo diodes, with a detection efficiency of around 50\% both at $532$~nm and $810$~nm.

The conversion efficiency of this setup for Gaussian beams is theoretically given by~\cite{albota2004esp}: 
\begin{eqnarray}
\eta=\sin^2(\pi/2 \sqrt{P_p/P_{max}}),\nonumber \\ \label{eq:efficiency}
P_{max}=\frac{c \epsilon_0 n_1 n_2 \lambda_1 \lambda_2 \lambda_p}{128 d^2_{\textrm{eff}} L h_m},
\end{eqnarray}
with pump beam power $P_p$, input and output wavelengths $\lambda_{1,2}$, the corresponding crystal refraction indices $n_{1,2}$, the effective nonlinearity $d_{\textrm{eff}}$, crystal length $L$ and the focussing factor $h_m$ (for Gaussian beams). The spot sizes of $\sim 70~\mu$m ($1550$) and $\sim 50~\mu$m ($810$), corresponding to a focussing parameter $\xi=\frac{L}{2z_R}$ (with Rayleigh length $z_R$) of about 0.8 for both beams, yield $h_m{\sim}0.6$. For the maximally available pump power of $1$~W and a single 4.3 mm long crystal we thus expect an efficiency of ${\sim}0.8\%$. Experimentally, calibration measurements with a $810$~nm laser diode resulted in $270$~nW of $532$~nm light converted from an input of $28~\mu$W at $810$~nm. Accounting for the wavelength difference and $\sim16\%$ optical loss, this implies an observed up-conversion efficiency of ${\sim}0.6\%$. The discrepancy between theory and measurement is likely due to a slightly lower effective non-linear coefficient caused by non-perfect periodic poling. 

For polarization-coherent operation we convert one photon of an entangled 810~nm/810~nm pair in the $\phi^{+}$ state created by our entangled photon source. From $7.3 \times 10^4$ counts per second (cps) input photon pairs, we detected $~15$ cps pairs after conversion, yielding an effective up-conversion efficiency of $\sim2\times10^{-4}$. Considering fiber coupling losses of $50$\% this corresponds to an intrinsic conversion efficiency---directly after the crystals---of about $0.04\%$. After accounting for the reduction by $50$\% because each crystal is pumped at half the pump power and another $\sim82\%$ because the beam focus is located between the two crystals instead of the crystal centers, this number is in good agreement to the theoretical efficiency---primarily limited by available pump power---and our auxiliary laser diode measurements.

A stringent way to demonstrate that polarization entanglement is preserved in the conversion process is the violation of a Bell inequality \cite{bell1964epr}, in our case the CHSH inequality \cite{clauser1969pet} for the converted, $532$~nm/$810$~nm polarization state:
\begin{equation} 
S=E(\alpha,\beta)-E(\alpha,\beta ')+ E(\alpha ' ,\beta)+ E(\alpha ',\beta ')\leq 2,
\end{equation}
where $E(\cdot,\cdot)$ are the correlations for joint polarization measurements on two photons along the angles $\alpha = 0^\circ, \alpha '= 45^\circ, \beta = 22.5^\circ, \beta '= 67.5^\circ $. A Bell value above the classical bound of $2$ implies that the measured state is incompatible with a local realistic model \cite{bell1964epr,clauser1969pet} and is thus entangled. 
With about 15 cps coincidence rate and integrating over 100 seconds for each measurement, we recorded the coincidences for the 16 necessary combinations of measurement angles. We obtained an experimental Bell parameter of 
\begin{equation}
S_{\textrm{exp}}=2.615\pm0.027,
\end{equation}
which violates the classical bound by more than $20$ standard deviations. The observation of entanglement in the output state is striking, because the original $810$~nm photon has been annihilated and created again in the $532$~nm mode---a rather invasive interaction.

\begin{figure}
\includegraphics[width=0.8\columnwidth]{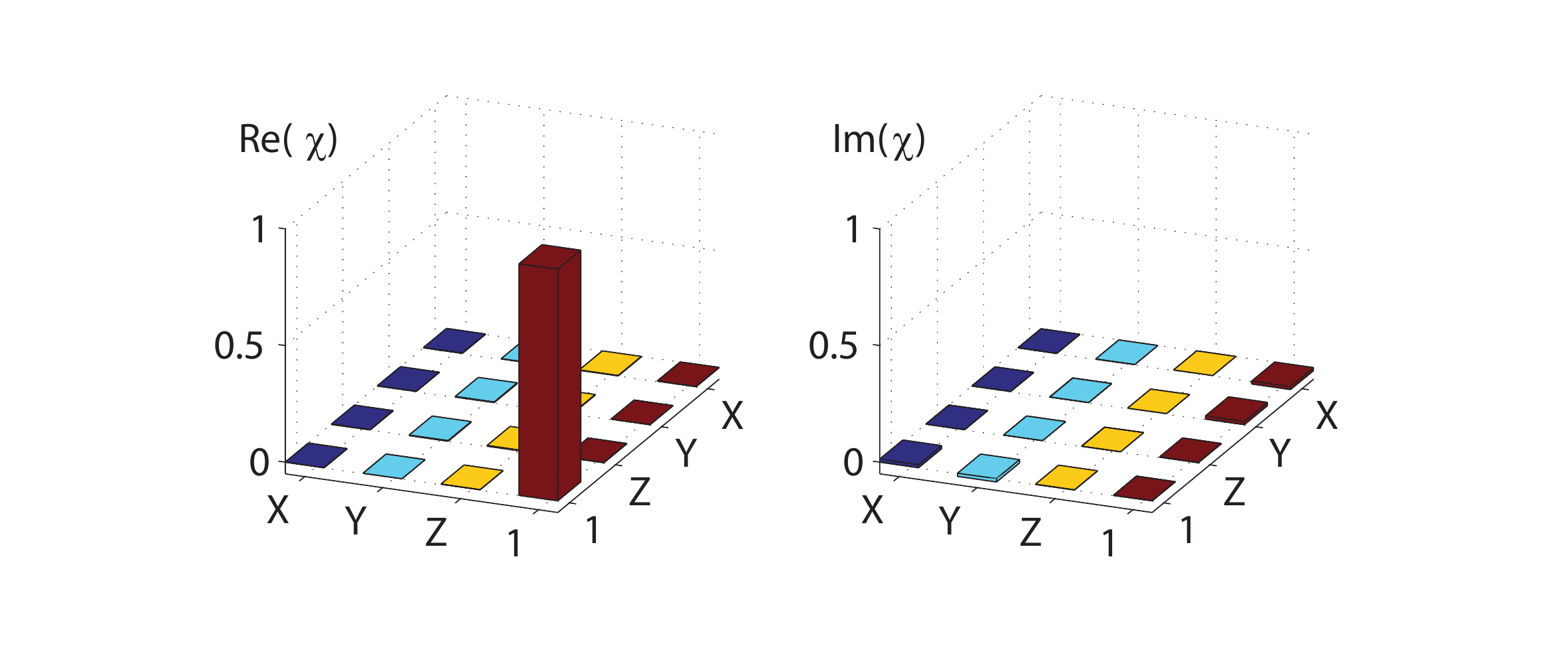}
\caption{Process matrix $\chi$ (Pauli-basis representation) for polarization-coherent up-conversion, characterised with an attenuated $810$~nm diode laser. The different elements of $\chi$ access what kind of operation---decomposed into the Pauli-operations---an input state is subject to, with the dominating element denoting the identity operation. The calculated process fidelity $\mathcal{F}$ and purity $\mathcal(P)$ are $\mathcal{F}=99.23\pm 0.01\%$ and $\mathcal{P}=98.54\pm0.01\%$. Error margins for $\mathcal{F}$  and $\mathcal(P)$ are determined assuming from Poissonian count statistics.}
\label{ptomo}
\end{figure}

To assess the quantum nature of the up-conversion process, we characterized it using tomographic techniques. We first performed process tomography \cite{poyatos1997ccq, chuang1997ped} with a strongly attenuated laser diode to assess the \emph{intrinsic}, i.e. independent of non-perfect detector-performance, dynamics of the entanglement transfer. We prepared the input states \{$\ket{H}, \ket{V}, \ket{D},\ket{A},\ket{R},\ket{L}$\} and, for each input, measured the the same set of 6 observables for the up-converted $532$~nm single-photon output, where $\ket{D}=(\ket{H}+\ket{V})/\sqrt{2}$, $\ket{A}=(\ket{H}-\ket{V})/\sqrt{2}$, $\ket{R}=(\ket{H}+i\ket{V})/\sqrt{2}$ and $\ket{L}=(\ket{H}-i\ket{V})/\sqrt{2}$. According to Eq.~\ref{eq:conversion}, for $\theta{=}0$ and balanced conversion, $\eta_{H}{=}\eta_{V}$, the ideal process matrix $\chi_{\textrm{ideal}}$ has a single non-zero element $(I,I)$ in the Pauli basis representation. This is very close to the reconstructed process matrix, figure~\ref{ptomo} which has a process fidelity~\cite{james2001mq} to the ideal case of $\mathcal{F}=99.23\pm0.01\%$. This indicates that the conversion process has excellent polarization coherence.

\begin{figure}[h!]
\includegraphics[width=.8\columnwidth]{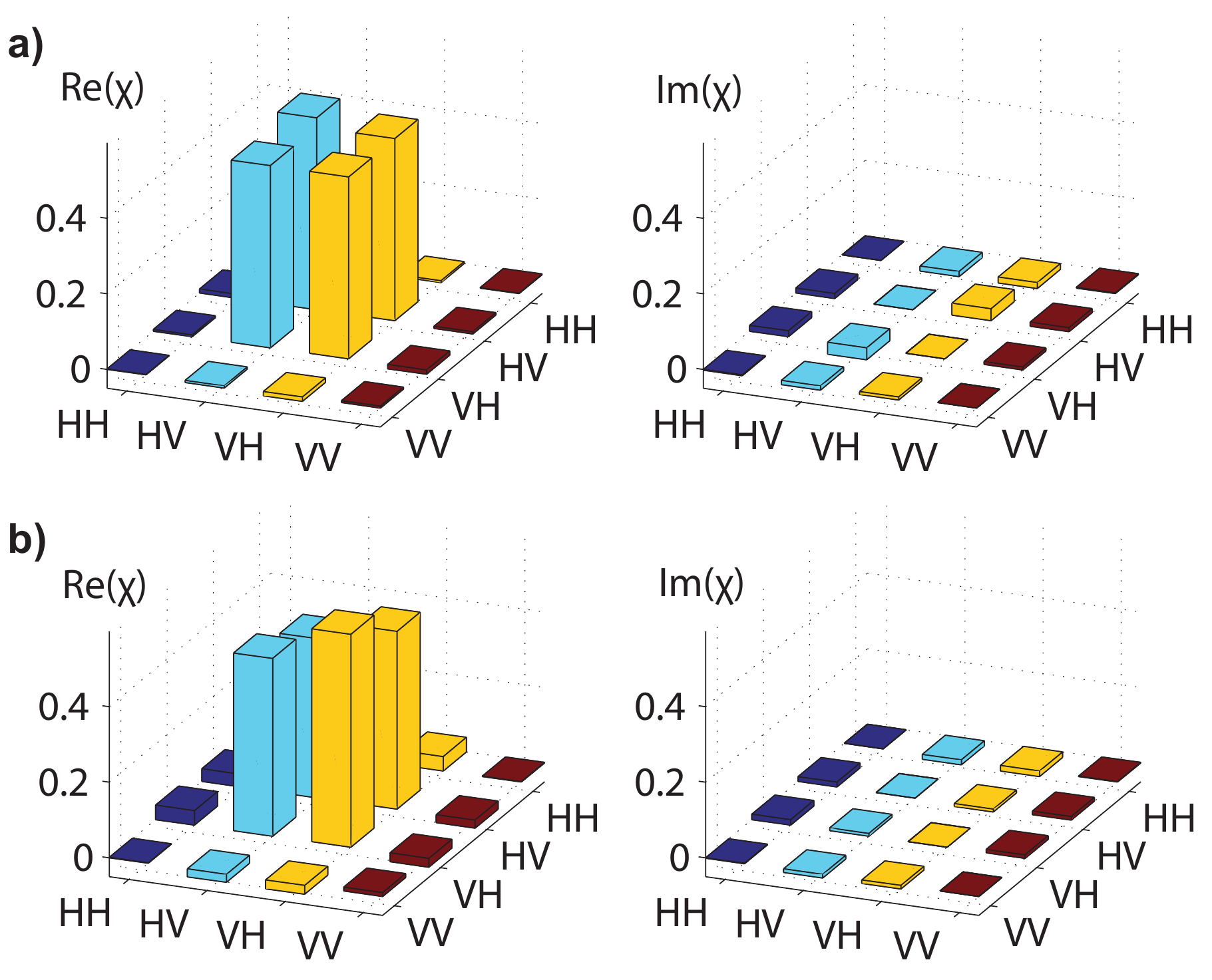}
\caption{Reconstructed (accidental-corrected) two-qubit density matrices of entangled input states and output states a) Input state ($810$~nm/$810$~nm) with a corresponding fidelity with respect to the $\psi^+$ state of $F_{\rm in}=97.93 \pm 0.03\%$ b) Output state ($810$~nm/$532$~nm) with $F_{\rm out}=96.7 \pm 0.2\%$. The fidelities as well as the values for the purities (P) and tangles (T) of the input and output states are summarized in figure~\ref{fidelities}. Error margins follow from Monte-Carlo simulations assuming errors from Poissonian count statistics.}
\label{rhoinout}
\end{figure}

We subsequently characterized in detail the entanglement transfer: we performed two-qubit quantum state tomography~\cite{james2001mq} on both the entangled photon input state and the entangled photon output state and compared the two (see figure~\ref{rhoinout}). For this we measured a total of $36$ combinations of the $6$ single-qubit observables \{$\ket{H},\ket{V},\ket{D}, \ket{A},\ket{R},\ket{L}$\}, for a measurement time of $1$ second for the $810~$nm$/810~$nm polarization-entangled input state and $100$ seconds for the $810~$nm$/532~$nm output state in which the first 810 nm photon remained unchanged. We used maximum-likelihood optimization to reconstructed the two-qubit density matrices from these measurements, and calculated several key diagnostic parameters: the input state fidelity (with the maximally entangled Bell state $\phi^{+}$) is $\mathcal{F}_{\rho_{\rm in}}=95.91\pm0.04\%$ and tangle is $\mathcal{T}_{\rho_{\rm in}}=84.3 \pm 0.1\%$. These values decrease to $\mathcal{F}_{\rho_{\rm out}}=93.8\pm0.3\%$ and $\mathcal{T}_{\rho_{\rm out}}=77 \pm 1\%$ for the (partially) up-converted states.

\begin{figure}[h!]
\includegraphics[width=0.45\columnwidth]{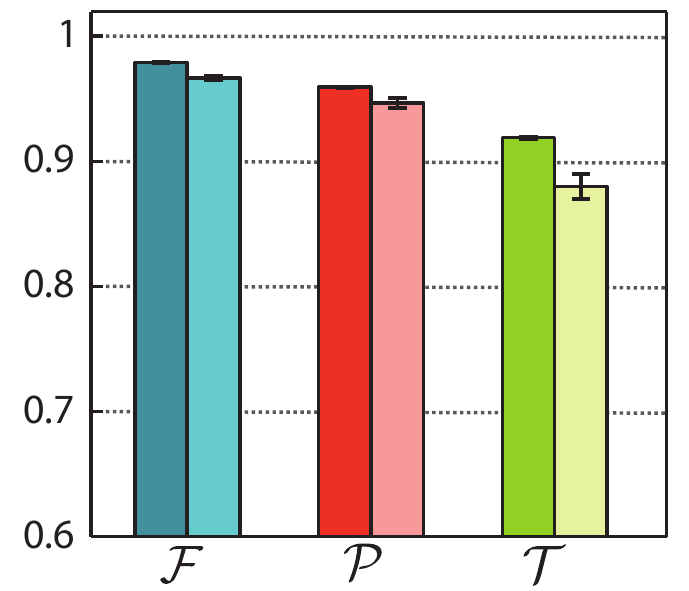}
\caption{Summary of the quality parameters for the accidental corrected  input states $\rho_{\rm in}$ (dark) and output states $\rho_{\rm out}$ (light): Fidelities $\mathcal{F}$, purities $\mathcal{P}$ and tangle $\mathcal{T}$. Error bars were obtained from Monte-Carlo runs of the tomographic reconstruction with assumed Poissonian count statistics.}
\label{fidelities}
\end{figure}

An error analysis shows that the most significant error contribution was caused by accidental coincidence counts, which occur when two photons from unrelated pairs are recorded within the coincidence time window. Double-pair emissions were negligible and we did not observe any statistically significant pump-induced back-ground counts. We estimated the accidental coincidence rates for every input and output measurement configuration by splitting one of the detector signals and measuring the coincidences with a relative time delay between the channels. We now subtract these accidentals from the raw coincidence counts in reconstructing our output states to probe the intrinsic quality of the up-conversion process. The resulting density matrices are shown in figure~\ref{rhoinout}. The parameters for the corrected output states are $\mathcal{F}_{\rho_{\rm out}}=96.7{\pm}0.2\%$, $\mathcal{P}_{\psi_{\rm out}}=94.7{\pm}0.4\%$ and $\mathcal{T}_{\rho_{\rm out}}=88{\pm}1\%$. These and the corresponding values calculated from the accidental-corrected (as explained above) input state are summarized in figure~\ref{fidelities}. The exceptional quality of the polarization entanglement transfer is further highlighted by the overlap fidelity between the entangled input and output states of $\mathcal{F}_{\rho_{\rm in},\rho_{\rm out}}=97.8\pm0.4\%$.

To conclude, we have shown and verified the conversion of polarization entanglement with intrinsically near unity fidelity using quantum state and process tomography. We furthermore violated a Bell inequality for the converted state. Our setup is flexible, compact and robust; it uses simple bulk nonlinear materials, requires no cryogenic or vacuum apparatus and is compatible with standard integrated-fibre and waveguide technologies. It is thus well suited for large-scale deployment in quantum networks and other quantum technologies where wavelength conversion is essential. Our conversion efficiencies are close to the theoretically calculated limit imposed by the available pump power and can be straight-forwardly enhanced by known techniques to achieve near unity single-photon efficiencies~\cite{albota2004esp, vandevender2004hes, langrock2005hes, thew2006lju}.  Specifically, for our polarization coherent design the efficiency can be significantly increased by using longer crystals and moving to bidirectionally pumped schemes (e.g.~Sagnac-, or Michelson-type interferometers~\cite{albota_polarization-independent_2006}). Importantly, with pump schemes like ours where the pump has a lower frequency then the converted photons~\cite{dong2008esp, kamada_efficient_2008} the conversion can remain free of pump-induced noise even at the required high pump power.

Converting $810$~nm to $532$~nm, as demonstrated here, can be important for various reasons; for example custom $532$~nm single-photon detectors can have up to $10$~times lower timing jitter than their $810$~nm counterparts and superconducting nanowire detectors as well as CCD-based imaging systems are more efficient at shorter wavelengths. Moreover, the wavelengths in our setup are interchangeable. Up-converting $1550$~nm photons can be achieved by pumping with $810$~nm, where powerful lasers are readily available.  Coherent frequency conversion also opens up avenues in fundamental physics, such as enabling access to superposition bases for color qubits \cite{ramelow2009dtc}. Finally, a future interesting challenge will be to also change the photons' spectral bandwidth during frequency conversion via suitably designed phase matching similar to~\cite{kielpinski_quantum_2011}. This could prove useful for interfacing photons with narrow-bandwidth quantum memories.   

During preparation of this manuscript we became aware of interesting related work involving entanglement transfer using four-wave-mixing in Rb vapor\cite{dudin_entanglement_2010}.

We thank M.~Hentschel and T.~Jennewein for valuable discussions. This  work  has  been  supported  by the Austrian Science Fund within SFB 015 P06, P20 and CoQuS (W1210), by the  ERC (Advanced Grant QIT4QAD) and the European Commission project QESSENCE.

\end{document}